# Magnetic Fields of Slowly Pulsating B Stars and β Cep Variables:

# Comparing Results from FORS1/2 and ESPaDOnS

Matt Shultz[a,b], Gregg Wade[b,a], Stefano Bagnulo[c], John Landstreet[c,d], Jason Grunhut[a,b], James Silvester[a,b], David Hanes[a] & the MiMeS Collaboration

[a]Queen's University, Kingston, ON
[b]Royal Military College of Canada, Kingston, ON
[c]Armagh Observatory, Armagh, Northern Ireland
[d]University of Western Ontario, London, ON

**Abstract.** Recently announced magnetic models for four SPB and β Cep stars, along with magnetic detections for two additional stars, have potentially doubled the number of known magnetic SPB and β Cep stars (see Grunhut *et al*., these proceedings). We have reanalyzed the published data and re-reduced archival low resolution spectropolarimetry collected with the FORS1/2 instruments at VLT on which the models were based, and compare them with high resolution data from the ESPaDOnS spectropolarimeter at CFHT, investigating previously noted inconsistencies between results from the two instruments.

## INTRODUCTION

Six stars were reported as magnetic by Hubrig *et al*., (2011) using FORS1/2 observations (one, $\xi^1$ CMa, was previously reported as magnetic by Hubrig *et al*. (2009) and confirmed with ESPaDOnS observations by Silvester *et al*., 2009). For four stars (α Pyx, B1.5III, βCep; 15CMa, B1Ib, β Cep; 33 Eri, B5V, SPB, SB1; $\xi^1$ CMa, B1III, β Cep), an inclination of the rotational axis from the line of sight *i*, magnetic obliquity β, dipolar field strength $B_d$, and projected rotational velocity $v\sin i$ were determined.

Models for three of those stars, together with the original published longitudinal field ($B_Z$) measurements, are shown in Figure 1. No models were derived for the remaining two stars (ε Lupi, B2IV-V, β Cep, SB; HY Vel, B3IV, SPB, SB1).

For those stars for which detailed models were developed, three simple models were tested against the published $B_Z$ measurements: zero field, static (mean) field and variable field. In all cases the reduced $\chi^2$ indicated a better fit for the variable field model. However, the reduced $\chi^2$ of the variable model fits is remarkably low. Examination of the periodograms (Fig. 1) shows that in most cases the published periods are not uniquely well-fit solutions: numerous possibilities exist beneath the 3σ confidence limit.

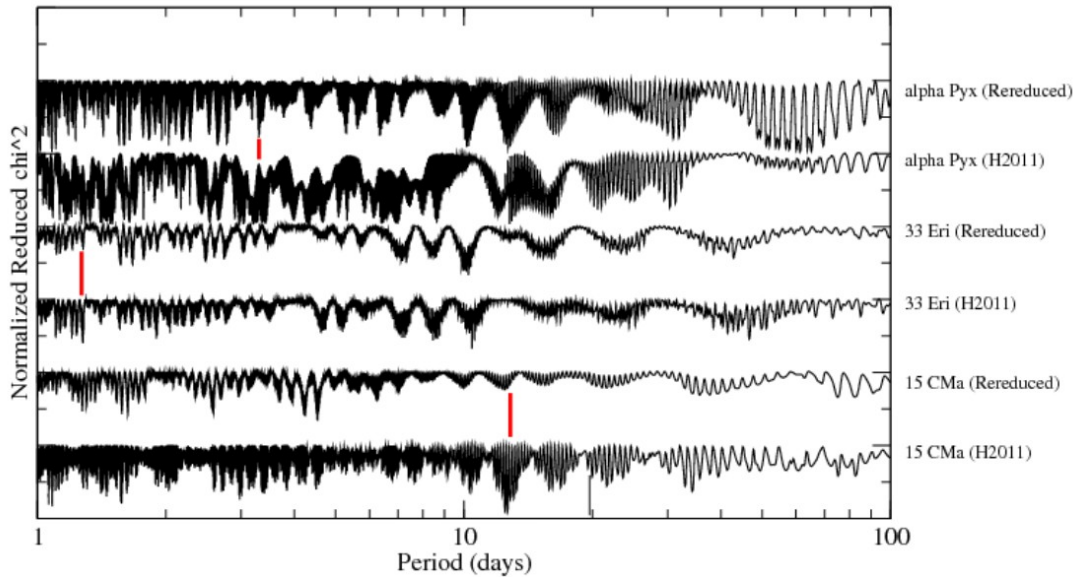

**FIGURE 1.** Sample periodograms of three of the stars modeled in H2011, generated from $B_Z$ as measured from the full spectrum original FORS data and the redreduced archival data. Model periods are indicated with short thick lines.

## COMPARISON TO ESPADONS OBSERVATIONS

In contrast to low-resolution techniques, high-resolution spectropolarimetry can resolve the metal lines of the spectrum. Using multiline analysis techniques such as Least Squares Deconvolution (LSD; Donati *et al.*, 1997), the polarization information from each line can be combined in order to achieve a much higher signal-to-noise ratio in the Stokes V profile than is available from any one line, enabling the detection and diagnosis of relatively weak magnetic fields in stars through modeling of the Stokes V LSD profiles as well as measurement of $B_Z$. In this case each profile is created from ~250 metal lines.

LSD profiles for three of the program stars for which detailed models were derived based on FORS data are shown in Fig. 3. No Zeeman signature is apparent in the LSD profiles of these targets. The resulting $B_Z$ measurements are shown in Fig. 2, phased according to the ephemerides published by Hubrig *et al.* (2011). In all cases $B_Z$ as determined with ESPaDOnS data is consistent with 0 G.

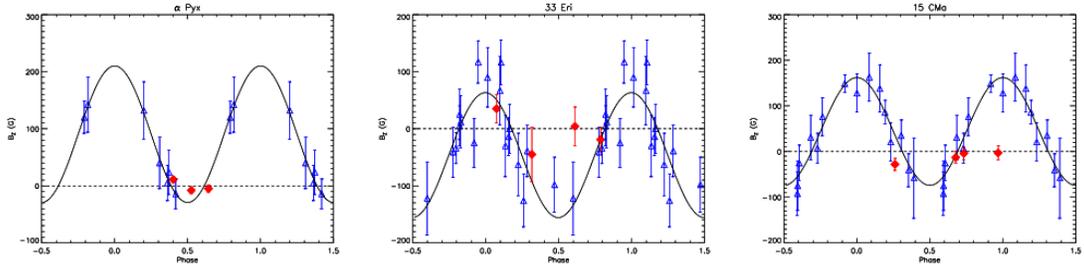

**FIGURE 2.** Comparison between $B_Z$ as measured with FORS1/2 (open triangles) and ESPaDOnS (filled circles). Left – right: α Pyx, 33 Eri, 15 CMa.

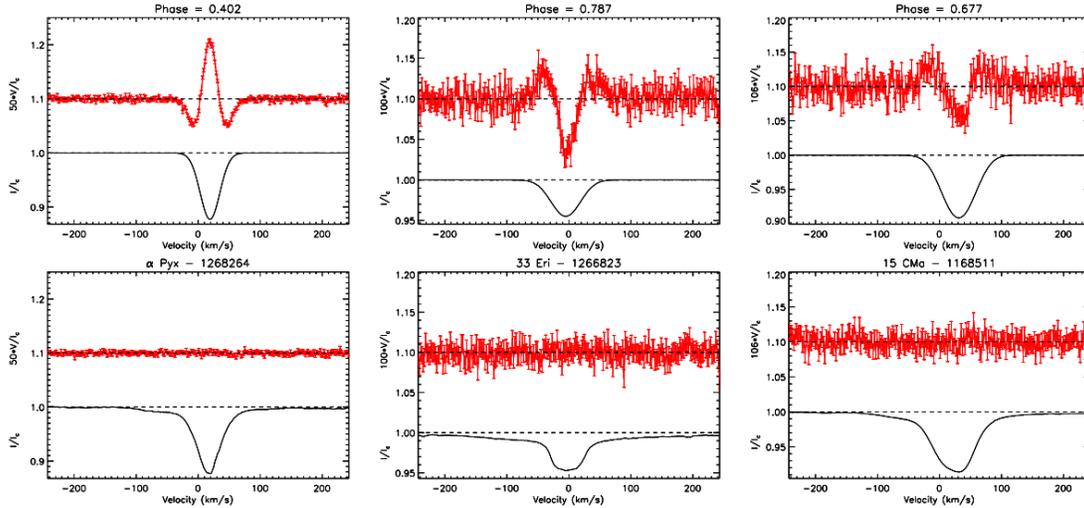

**FIGURE 3.** Above: models; below: observed; Left – right: α Pyx, 33 Eri, 15 CMa.

Configurations exist for which $B_Z = 0$ G, but a Zeeman signature can still be detected. Synthetic LSD profiles generated using the parameters (i, β, $v\sin i$, $B_d$) of the models by Hubrig *et al*. are shown in Fig. 3, with Gaussian noise statistics and error bars extracted from the corresponding observational LSD profiles. In all cases the models predict Zeeman signatures that should result in unambiguous detections, which are not observed.

## COMPARISON TO RE-REDUCED FORS1/2 DATA

Archival data was re-reduced using a new pipeline. In addition to $B_Z$ measurements based on H Balmer lines and on the full spectrum, measurements based on only the metal lines were extracted, as well as diagnostic nulls. All measurements were compared to the same simple models as for the published data. The rereduced $B_Z$ measurements are shown as function of the original $B_Z$ measurements in Fig. 4.

For ε Lupi the mean $B_Z$ is significant to 5.8σ at 77±12 G, with agreement between both reductions and the results from ESPaDOnS using LSD. For α Pyx, in only one case is a re-reduced measurement formally inconsistent with a null result. Curiously it is both larger than the original, and of opposite polarity. Investigation revealed anomalies in the Stokes V spectrum: the best-resolved Balmer lines show no sign of a Zeeman signature, while the metal line measurements are entirely consistent with the zero field model. For 33 Eri, most measurements are formally consistent with a null result, and the mean $B_Z$ is significant to only 1.5σ. For 15 CMa, most measurements were formally consistent with a null result, and there is noticeable scatter in $B_Z$ between reductions (see Fig. 4). For HY Vel, the mean $B_Z$ from Stokes V is significant to 3σ, however, there is only a small difference between Stokes V and N, and the metal lines are indistinguishable between V and N. Note that for $ξ^1$ CMa, a known magnetic star, there is close agreement between $B_Z$ between reductions.

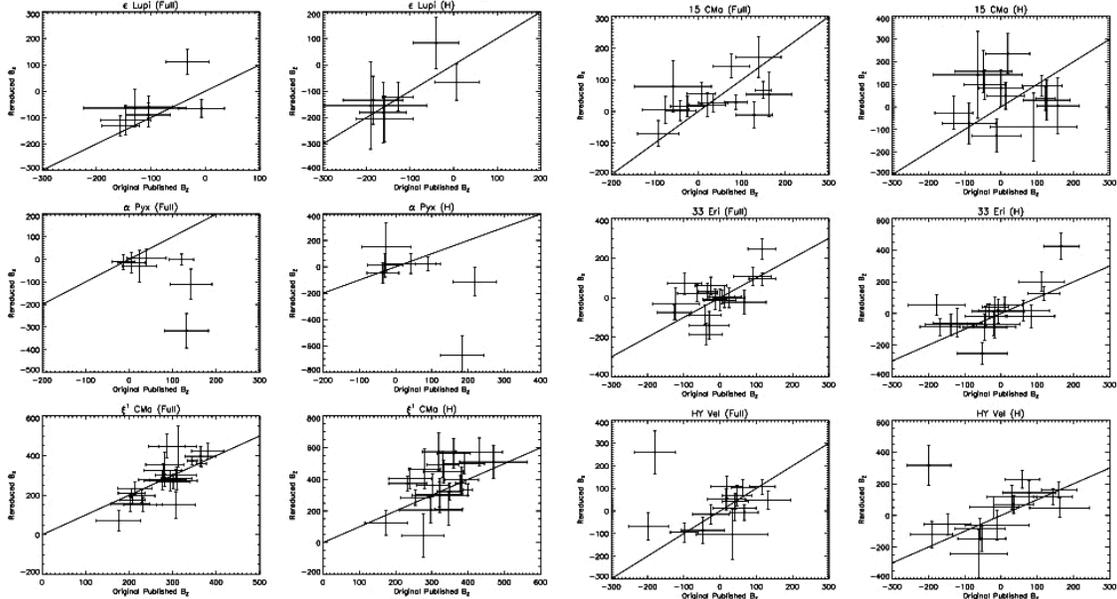

**FIGURE 3.** $B_Z$ from the re-reduced observations as a function of $B_Z$ from the original observations. The solid line indicates x=y.

## CONCLUSIONS AND FUTURE WORK

Magnetic fields are not confirmed in either ESPaDOnS or the reanalyzed FORS1/2 data for four of the six stars observed. For three stars modeled in detail by Hubrig et al. (α Pyx, 15 CMa, 33 Eri) no magnetic field is detected, and differences between the original and re-reduced results suggest the proposed field geometries and periods are artifacts of the data analysis and reduction procedures. The fourth star modeled in detail by Hubrig *et al.* ($ξ^1$ CMa) was previously detected as magnetic by both FORS1/2 and ESPaDOnS; an alternate model based on ESPaDOnS observations will be presented by Wade *et al.* (in preparation.) For one star reported as magnetic by Hubrig *et al.* (ε Lupi) a magnetic field is confirmed and further ESPaDOnS observations are being collected; for one star reported as magnetic by Hubrig *et al.* (HY Vel) no field is detected, although in this case poor observing conditions limited the sensitivity and further observations are planned.

## ACKNOWLEDGMENTS

The National Science and Engineering Research Council, Veronique Petit, James Silvester, Huib Henrichs.

# REFERENCES

1. Donati, J.-F., Semel, M., Carter, B. D., Rees, D. E. & Collier Cameron, A., *MNRAS*, **291**, 658, (1997)
2. Hubrig, S., Briquet, M.,De Cat, P., Schöller, M., Morel, T. & Ilyin, I., *AN*, **330**, 317 (2009)
3. Hubrig, S.; Ilyin, I.; Schöller, M.; Briquet, M.; Morel, T. & De Cat, P., *ApJ*, **726**, 5 (2011)
4. Silvester, J., Neiner, C., Henrichs, H. F., Wade, G. A., Petit, V., Alecian, E., Huat, A.-L., Martayan, C., Power, J. & Thizy, O., *MNRAS*, **398**, 1505 (2009)